\documentclass[11pt]{JHEP3}
\usepackage{amssymb}
\usepackage{amsmath}
\usepackage{amsfonts}
\usepackage{epsfig}

\author{Vyacheslav S. Rychkov \\
\ \\
Scuola Normale Superiore and INFN, Piazza dei Cavalieri 7, 56126
Pisa, Italy}
\title{D1-D5 black hole microstate counting\\ from supergravity}

\abstract{We quantize the moduli space of regular D1-D5
microstates, directly from Type IIB SUGRA. The moduli space is
parametrized by a smooth closed non-selfintersecting curve in four
dimensions, and we derive that the components of the curve satisfy
chiral boson commutation relations, with the correct value of the
effective Planck constant previously conjectured using U-duality.
We use the Crnkovi\'c-Witten-Zuckerman covariant quantization
method, previously used to quantize the `bubbling AdS' geometries,
combined with a certain new `consistency condition' which allows
us to reduce the computation to quantizing perturbations around
the plane wave.}

\begin{document}

\section{Introduction}

Black hole entropy can often be derived in string theory by counting excited
states of equivalent D-brane configurations. The ensemble of D-brane states
with fixed charges turns into a classical black hole when the Newton constant
is increased. The main point of Mathur's `fuzzball' idea \cite{Mathur} is to
ask what happens with the individual (pure) D-brane states in the same
process. AdS/CFT suggests that these may become smooth horizonless geometries.
This opens up an exciting possibility: if these `black hole microstate
geometries' can be identified and counted, then the black hole entropy can be
reproduced directly from supergravity. The goal of this paper is to do exactly
this in the 2-charge black hole case, for which the microstate geometries are
relatively well understood.

The plan of the paper is as follows. In Section 2 we collect the necessary
information about the D1-D5 black hole and its microstates, and formulate our
main result. In Section 3 we review the quantization method based on
evaluating the symplectic form. In addition to our previous techniques
\cite{MR,5auth}, we derive a `consistency condition' which is a strong
constraint on the form of the restricted symplectic form. In the present case,
it can be used to predict the symplectic form up to a prefactor. To complete
the calculation, it is sufficient to evaluate the prefactor for perturbations
around a conveniently chosen specific spacetime, which we do in Section 4. We
conclude in Section 5 with a general discussion of the future of Mathur's program.

\textbf{Note added.} When this paper was being prepared for publication, we
received an interesting paper \cite{Jevicki} where, among other results, the
`planar curve' part of moduli space was quantized directly (using the method
of \cite{MR}). Our result agrees with \cite{Jevicki}, while our new method
based on the consistency condition makes the computation somewhat simpler.

\section{D1-D5 geometries}

\subsection{Fields and charges}

The D1-D5 black hole microstate geometries are regular solutions of Type
IIB\ SUGRA. The nontrivial fields (the metric, the dilaton, and the RR
two-form) are given by \cite{LM,LMM}%
\begin{align}
ds^{2}  &  =e^{-\Phi/2}ds_{\text{string}}^{2}\,,\nonumber\\
ds_{\text{string}}^{2}  &  =\frac{1}{\sqrt{f_{1}f_{5}}}\left[  -(dt+A)^{2}%
+(dy+B)^{2}\right]  +\sqrt{f_{1}f_{5}}\,d\mathbf{x}^{2}+\sqrt{f_{1}/f_{5}%
}\,d\mathbf{z}^{2}\,,\nonumber\\
e^{2\Phi}  &  =f_{1}/f_{5}\,,\nonumber\\
C  &  =\frac{1}{f_{1}}\left(  dt+A\right)  \wedge\left(  dy+B\right)
+\mathcal{C}\,,\nonumber\\
dB  &  =\ast_{4}dA\,,\qquad d\mathcal{C}=-\ast_{4}df_{5}\,,\label{micro}\\
f_{5}  &  =1+\frac{Q_{5}}{L}\int_{0}^{L}\frac{ds}{|\mathbf{x}-\mathbf{F}%
(s)|^{2}}\,,\nonumber\\
f_{1}  &  =1+\frac{Q_{5}}{L}\int_{0}^{L}\frac{|\mathbf{F}^{\prime}(s)|^{2}%
ds}{|\mathbf{x}-\mathbf{F}(s)|^{2}}\,,\nonumber\\
A  &  =\frac{Q_{5}}{L}\int_{0}^{L}\frac{F_{i}(s)\,ds}{|\mathbf{x}%
-\mathbf{F}(s)|^{2}}\,dx^{i}\,.\nonumber
\end{align}
The solutions are asymptotically $M^{5}\times S^{1}\times T^{4}$; $y$ and
$\mathbf{z}$ denote the $S^{1}$ and $T^{4}$ directions. The moduli space is
parametrized by a closed curve%
\begin{equation}
x_{i}=F_{i}(s)\quad(0<s<L,i=1\ldots4),
\end{equation}
which is assumed to be smooth and non-selfintersecting. Its parameter length
has to satisfy%
\begin{equation}
L=\frac{2\pi Q_{5}}{R},
\end{equation}
where $R$ is the coordinate radius of $S^{1}$. Under these
conditions the above geometries are completely regular. It should
be noted that this description is somewhat redundant, since a
constant shift of parameter $s\to s+h$ would produce the same
geometry; this redundancy will play a role below.

In Mathur's approach to black hole microstates \cite{Mathur}, solutions
(\ref{micro}) are supposed to represent microstates of the spherically
symmetric 2-charge geometry, which is given by Eqs.~(\ref{micro}) if we
replace
\begin{equation}
f_{1,5}\rightarrow1+\frac{Q_{1,5}}{|x|^{2}},\quad A_{i},B_{i}\rightarrow0\,.
\end{equation}
This system is equivalent to $M^{5}\times S^{1}\times T^{4}$ with $N_{1}$
D1-branes wrapping $S^{1}$ and $N_{5}$ D5 branes wrapping the full
$S^{1}\times T^{4}$. The brane numbers are related to the charges by%
\begin{equation}
Q_{5}=g_{s}N_{5},\quad Q_{1}=\frac{g_{s}}{V_{4}}N_{1}%
\end{equation}
(in the units $\alpha^{\prime}=1$), where $g_{s}$ is the string
coupling and $(2\pi)^{4}V_{4}$ is the coordinate volume of
$T^{4}$. The D1-D5 system is known to have the macroscopic
entropy\footnote{See \cite{Marolf} for a derivation of this
entropy by using the Dirac-Born-Infeld effective action to
quantize supertubes.}
\begin{equation}
S=2\pi\sqrt{2N_{1}N_{5}}\text{.} \label{entropy}%
\end{equation}
This entropy was not yet reproduced as a thermodynamic Bekenstein-Hawking
entropy, because the spherically symmetric geometry has a classical horizon of
zero area. A nonzero horizon is expected to appear once higher-curvature
corrections to the two-derivative Type IIB SUGRA are included, similar to how
it happens in the heterotic compactifications \cite{Sen}.

\subsection{Counting D1-D5 microstates}

In this paper we will reproduce (a finite fraction of) the entropy
(\ref{entropy}) by counting microstate geometries (\ref{micro}). To do this,
we will have to quantize the curve $\mathbf{F}(s)$ (see footnote 3 in
\cite{LMM}). As we will show, in quantum theory $\mathbf{F}(s)$ acquires
commutation relations:%
\begin{equation}
\left[  F_{i}(s),F_{j}^{\prime}(\tilde{s})\right]  =i\pi\mu^{2}\delta
_{ij}\delta(s-\tilde{s})\,. \label{commutator}%
\end{equation}
This commutation relation has been previously conjectured by using
the fact that the D1-D5 system can be U-dualized into the FP
system (the multiply wound fundamental string carrying momentum
along a compactified direction). The curve $\mathbf{F}(s)$ in the
D1-D5 picture was identified with the profile of a chiral
excitation of the dual string according to a simple
proportionality relation%
\begin{equation}
\mathbf{F}(s)=\mu\mathbf{F}_{\text{FP}}(s),\qquad\mu=\frac{g_{s}}{R\sqrt
{V_{4}}}\,.
\end{equation}
The fundamental string is quantized using the quadratic action
\begin{equation}
\frac{1}{4\pi}\int(\dot{X}_{i}^{2}-X_{i}^{\prime2})d\tau d\sigma\,,
\end{equation}
and thus satisfies the commutation relation%
\begin{equation}
\lbrack X_{i}(\sigma),\dot{X}_{i}(\tilde{\sigma})]=i2\pi\delta(\sigma
-\tilde{\sigma})\delta_{ij}\,.
\end{equation}
The commutator of its chiral component is $1/2$ of that:%
\begin{equation}
\lbrack F_{\text{FP}i}(s),F_{\text{FP}j}^{\prime}(\tilde{s})]=i\pi\delta
_{ij}\delta(s-\tilde{s})\,,
\end{equation}
and (\ref{commutator}) follows.

Later in this paper we will show how, instead of such duality-based reasoning,
one can derive (\ref{commutator}) directly from the SUGRA action. At this
point let us demonstrate how this result can be used to compute the degeneracy
of the D1-D5 system. From the SUGRA point of view, we must count the number of
microstate geometries with fixed charges $Q_{1,5}$, which translates into
counting the number of curves $\mathbf{F}(s)$ satisfying the relation%
\begin{equation}
Q_{1}=\frac{Q_{5}}{L}\int_{0}^{L}|\mathbf{F}^{\prime}(s)|^{2}ds\,.
\label{Q1Q5}%
\end{equation}
Classically, this question cannot be answered --- there are infinitely many
such curves. However, once we take the commutator (\ref{commutator}) into
account, we should expand $\mathbf{F}(s)$ into quantum oscillators:
\begin{align}
&  \mathbf{F}(s)=\mu\sum_{k=1}^{\infty}\frac{1}{\sqrt{2k}}\left(
\mathbf{c}_{k}e^{i\frac{2\pi k}{L}s}+\mathbf{c}_{k}^{\dagger}e^{-i\frac{2\pi
k}{L}s}\right)  \,,\nonumber\\
&  [c_{k}^{i},c_{k^{\prime}}^{j\dagger}]=\delta^{ij}\delta_{kk^{\prime}}\,,\\
&  \left\langle \int_{0}^{L}:|\mathbf{F}^{\prime}(s)|^{2}:ds\right\rangle
=\frac{\left(  2\pi\right)  ^{2}}{L}\mu^{2}N_{\text{osc}}\,,\nonumber\\
&  N_{\text{osc}}=\sum_{k=1}^{\infty}k\langle\mathbf{c}_{k}^{\dagger
}\mathbf{c}_{k}\rangle\,,\nonumber
\end{align}
In such a quantum theory relation (\ref{Q1Q5}) takes the form
\begin{equation}
N_{1}N_{5}=N_{\text{osc}}\,.
\end{equation}
Thus the degeneracy of states is equal to the degeneracy of the $N_{1}N_{5}$
energy level in the system of 4 chiral bosons:%
\begin{equation}
\Gamma\sim\exp\left(  2\pi\sqrt{\frac{c}{6}N_{1}N_{5}}\right)  ,\quad c=4\,.
\end{equation}
It follows that the microstate geometries (\ref{micro}) account
for a finite fraction of the full D1-D5 entropy. It is well known
that they are insufficient to recover the full entropy: one would
have to consider solutions corresponding to the string vibrating
in the $T^{4}$ directions \cite{LMM} and to the fermionic
excitations of the string (see \cite{Marika}).

\section{Quantization from SUGRA}

\subsection{Symplectic form quantization}

We would now like to derive the commutation relation (\ref{commutator})
directly from SUGRA. There are several reasons why such a result would be
welcome. First, it is important to know as a matter of principle that the
commutators can be extracted from the classical geometries without any further
input. Second, the U-duality-based derivation of (\ref{commutator}) described
above is not fully satisfactory: 1) it may not be easily generalizable to
3-charge microstate geometries which seem to have a very complicated moduli
space parametrized by 4-dimensional hyper-Kaehler metrics \cite{BW}; 2)
strictly speaking, the shape of the curve is not guaranteed to be a
duality-invariant notion (only the degeneracies of states are).

Notice that we expect nontrivial commutation relations (\ref{commutator}) to
appear already \textit{among the functions parametrizing the moduli space}.
This situation should be contrasted with the case of Manton's soliton
scattering, where positions of solitons by themselves commute, and to get
nontrivial quantization one has to augment the phase space by adding momenta,
corresponding to slow soliton motion. This difference can be traced to the
stationary, as opposed to static, nature of our solutions.

The standard approach to quantization would be to work with the quadratic
action for small fluctuations around every point in the moduli space. However,
this is not technically feasible, since for a general geometry (\ref{micro})
there does not seem to be a convenient basis at hand into which to expand
these small fluctuations. Another problem is that, typically, the quadratic
action would couple fluctuations along moduli space to fluctuations orthogonal
to it. This indicates that quantizing the quadratic action is a more difficult
problem than the one we have to solve.

All these difficulties were explained in \cite{MR,5auth}, where it was
proposed that instead of the quadratic action, it is more efficient to think
directly in terms of the classical equivalent of the commutation relations ---
the Poisson brackets. In a general formulation of the problem we are given a
classical dynamical system with phase space coordinates having the standard
Poisson brackets%
\begin{equation}
\{q^{I},p^{J}\}=\delta^{IJ}\,.
\end{equation}
We are also given a subspace $\mathcal{M}$ of the full phase space
parametrized by some coordinates $x^{A}$:%
\begin{equation}
q=q(x),\quad p=p(x)\,.
\end{equation}
The problem is to find the induced Poisson brackets: $\{x^{A},x^{B}\}=?$ To
solve the problem let us consider the symplectic form of the theory:%
\begin{equation}
\Omega=dp^{I}\wedge dq^{I}\,.
\end{equation}
The symplectic structure on $\mathcal{M}$ is given by the pull-back (i.e.
restriction) of $\Omega$:
\begin{align}
\Omega|_{\mathcal{M}}  &  =\omega_{AB}(x)dx^{A}\wedge dx^{B}\,,\nonumber\\
\omega_{AB}  &  =\frac{\partial p^{I}}{\partial x^{[A}}\frac{\partial q^{I}%
}{\partial x^{B]}}\,.
\end{align}
The induced Poisson brackets are simply given by the inverse of $\omega_{AB}$:%
\begin{equation}
\{x^{A},x^{B}\}=\frac12\omega^{AB}\,.
\end{equation}
This argument shows why it is convenient to think in terms of the symplectic
form: it encodes the Poisson brackets in a covariant way.

In our particular situation the phase space will be that of Type IIB SUGRA,
the subspace $\mathcal{M}$ being the moduli space of D1-D5 geometries
parametrized by closed curves. The Einstein-frame action of Type IIB SUGRA in
the relevant sector is given by%
\begin{equation}
S_{\text{IIB}}=\frac{1}{(2\pi)^{7}g_{s}^{2}}\int\sqrt{-g}\left[  R-\frac{1}%
{2}(\partial\Phi)^{2}-\frac{1}{2}e^{\Phi}|F_{3}|^{2}\right]  ,\quad
F_{3}=dC\,. \label{IIB}%
\end{equation}
To define the phase space variables, we put the theory in the Hamiltonian
form. Dynamical degrees of freedom on a surface of constant time will be given
by the spatial components of the metric and of the two-form, as well as by the
value of the dilaton:%
\begin{equation}
q=\left\{  g_{ab},C_{ab},\Phi\right\}  \,.
\end{equation}
The remaining components of the metric and of the two-form ($g_{tt}%
,g_{ta},C_{ta}$) will appear in the action only as Lagrange multipliers,
i.e.~without time derivatives. The symplectic form of the theory is given by%
\begin{equation}
\Omega=\int_{t=const}d^{9}x\,\sum_{q}\delta\Pi_{q}(x,t)\wedge\delta
q(x,t),\quad\Pi_{q}=\frac{\partial L}{\partial\dot{q}}\,. \label{sform0}%
\end{equation}
Here $\delta$ denotes the differential in the space of fields, not to be
confused with the spacetime differential $dx^{\mu}$. This equation can be
rewritten in the Crnkovi\'{c}-Witten-Zuckerman \cite{CW,Z} covariant formalism
as an integral over a Cauchy surface $\Sigma$ of the symplectic current:%
\begin{align}
\Omega &  =\int d\Sigma_{\mu}\,J^{\mu},\label{cauchy}\\
J^{\mu}  &  =\delta\left(  \frac{\partial L}{\partial\partial_{\mu}\psi_{A}%
}\right)  \wedge\delta\psi_{A}\,. \label{current}%
\end{align}
Here $\psi_{A}$ runs over all the fields of the theory. If we choose
$\Sigma=\{t=const\}$, the contributions of Lagrange multipliers drop out, and
we recover (\ref{sform0}). The symplectic current $J^{\mu}$ has several useful
properties: 1) it is conserved as a consequence of the classical equations of
motion (and thus $\Omega$ is independent of the choice of the Cauchy surface),
it changes by a total derivative under a gauge transformation (and thus
$\Omega$ is gauge invariant). Moreover, both $J^{\mu}$ and $\Omega$ are
invariant under point transformations of the elementary fields.

The symplectic form of the relevant sector of Type IIB SUGRA can be written
as
\begin{align}
\Omega &  =\frac{1}{(2\pi)^{7}g_{s}^{2}}\int d\Sigma_{\mu}\,J^{\mu
}\,,\nonumber\\
J^{\mu} &  =J_{g}^{\mu}+J_{F}^{\mu}+J_{\Phi}^{\mu}\,.\label{formIIB}%
\end{align}
The three terms here are the gravity, two-form, and dilaton symplectic
currents, which can be computed using Eq.~(\ref{current}) from the
corresponding parts of the action (\ref{IIB}). We have%
\begin{align}
J_{g}^{\mu} &  =-\delta\Gamma_{\alpha\beta}^{\mu}\wedge\delta(\sqrt
{-g}g^{\alpha\beta})+\delta\Gamma_{\alpha\beta}^{\beta}\wedge\delta(\sqrt
{-g}g^{\mu\alpha})\,,\label{CW}\\
J_{F}^{\mu} &  =-\delta\left(  \sqrt{-g}e^{-\Phi}F^{\mu|\alpha\beta|}\right)
\wedge\delta C_{|\alpha\beta|}\,,\label{Fs}\\
J_{\Phi}^{\mu} &  =-\delta\left(  \sqrt{-g}\partial^{\mu}\Phi\right)
\wedge\delta\Phi\,.
\end{align}
The $J_{g}^{\mu}$ is known as the Crnkovi\'{c}-Witten current \cite{CW}. When
deriving it from the gravitational action, one should use the so-called
$\Gamma\Gamma-\Gamma\Gamma$ Lagrangian, which contains only the first
derivatives of the metric and differs from the standard Einstein-Hilbert
Lagrangian by a total derivative term.

\subsection{Consistency condition}

The above logic has been laid out in \cite{MR,5auth}, where it was used to
quantize the moduli space of the `bubbling AdS' geometries \cite{LLM}. It can
be used to compute the symplectic form on any subspace of the full phase space
of gravity. Now we will add a new ingredient to the discussion. Namely, we
observe that the subspace we are dealing with is not just \textit{any}
subspace: it is rather special in that it consists of time-independent
solutions. It turns out that in this case there is an additional piece of
information which can simplify the derivation of the restricted symplectic
form. In particular, this information will allow us to predict the resulting
symplectic form up to a coefficient, without doing any difficult computations.

In fact, the idea we are about to explain applies to any subspace
$\mathcal{M}$ which is invariant under the Hamiltonian evolution. Thus we
consider a general Hamiltonian system ($H,\Omega)$ with a Hamiltonian $H$ and
a symplectic form $\Omega$. Let us restrict $\Omega$ and $H$ to $\mathcal{M}$:%
\begin{equation}
\omega=\Omega|_{\mathcal{M}},\qquad h=H|_{\mathcal{M}}\,.
\end{equation}
Now we have two hamiltonian flows on $\mathcal{M}$: the original flow
($H,\Omega$), which leaves $\mathcal{M}$ invariant by assumption, as well as
the flow ($h,\omega$) generated by the restricted objects.

\textbf{Theorem (\textit{Consistency condition}):} These flows are equivalent
on $\mathcal{M}$: ($H$,$\Omega$)$\equiv(h,\omega)$.

The proof is very simple. The solutions of the Hamilton equations can be
obtained in the first-order formalism as stationary curves of the functional%
\begin{equation}
\int dt\left[  K_{i}(X)\dot{X}^{i}-H(X)\right]  \,,
\end{equation}
where $K$ is a one-form such that $\Omega=dK$. Restricting $\Omega$ to
$\mathcal{M}$ is equivalent to restricting $K$. A stationarity curve will of
course remain stationary under a smaller class of variations which do not take
the curve outside $\mathcal{M}$. It follows that any curve of the flow
($H,\Omega$) must satisfy equations of motion of $(h,\omega)$. Q.E.D.

To apply this theorem in practice, we must evaluate the on-shell Hamiltonian
(i.e.~the solution energy) $H|_{\mathcal{M}}$ as a functional on the moduli
space. From the form of this functional we get a \textit{consistency
condition} which should be satisfied by $\omega$: the Hamiltonian equations
derived from $H|_{\mathcal{M}}$ and $\omega$ should coincide with the
evolution on the moduli space implied by the form of the solutions.

The energy of the D1-D5 microstate geometries is given by%
\begin{equation}
H|_{\mathcal{M}}=\frac{RV_{4}}{g_{s}^{2}}\left(  \frac{Q_{5}}{L}\int_{0}%
^{L}|\mathbf{F}^{\prime}(s)|^{2}ds+Q_{5}\right)  \,. \label{energy}%
\end{equation}
It can be evaluated using the standard general relativity formula for the
asymptotically flat spacetimes%
\begin{equation}
H=\frac{1}{(2\pi)^{7}g_{s}^{2}}\int_{\partial\Sigma}\left(  \frac{\partial
g_{ab}}{\partial x^{b}}-\frac{\partial g_{bb}}{\partial x^{a}}\right)  n^{a}%
\end{equation}
where $\partial\Sigma$ is the asymptotic boundary, which in our case is
$S^{1}\times T^{4}$ times the 3-sphere at large $|x|,$ $g_{ab}$ is the spatial
metric (i.e.$~a,b$ run over the coordinates $\mathbf{x},y,\mathbf{z}),$ and
$n^{a}$ is the outside unit normal to $\Sigma$. Notice that (\ref{energy})
agrees with the total mass of the D-branes:%
\[
E_{\text{tot}}=g_{s}^{-1}(N_{1}R+N_{5}RV_{4})\,.
\]

The consistency condition implies that the Poisson brackets on the moduli
space should be such that the Hamilton equation%
\begin{equation}
\frac{dF_{i}}{dt}=\{F_{i},H|_{\mathcal{M}}\}
\end{equation}
be compatible with the time-independence of the microstate geometries. A
moment's thought shows that the only nontrivial allowed equation is%
\begin{equation}
\label{EOM}\frac{dF_{i}}{dt}=\text{const.}\frac{dF_{i}}{ds}\,,
\end{equation}
so that%
\begin{equation}
\label{shift}F_{i}(s,t)=F_{i}(s+\text{const.}t)\,.
\end{equation}
Such, and only such, dynamics leads to time-independent geometries, since the
metric and other fields are given in terms of contour integrals which remain
unchanged under the constant shift of parameter (\ref{shift}).

The Hamiltonian (\ref{energy}) leads to Eq.~(\ref{EOM}) if and only if the
Poisson brackets have the form:%
\begin{equation}
\{F_{i}(s),F_{j}^{\prime}(\tilde{s})\}=\alpha\,\delta_{ij}\delta(s-\tilde
{s})\,, \label{bracket}%
\end{equation}
where $\alpha$ is a constant. The corresponding symplectic form is
thus fixed up to a proportionality coefficient\footnote{Notice
that this symplectic form is invariant under the infinitesimal
transformation $\delta\mathbf{F}(s)\to \delta
\mathbf{F}(s)+\epsilon \mathbf{F}'(s)$. Thus it restricts nicely
to the `true' moduli space which is the space of curves modulo
constant parameter shifts.
In fact the whole above discussion could be phrased, completely equivalently, in terms of this `true' moduli space.}:%
\begin{equation}
\Omega=\frac{1}{2\alpha}\int\delta F_{i}^{\prime}(s)\wedge\delta
F_{i}(s)\,ds\,. \label{toshow}%
\end{equation}

Notice that the precise value of $\alpha$ cannot be established by this
argument --- it will have to be found by an explicit calculation. Notice also
that $\alpha$, although a constant for each $\mathbf{F}(s)$, could in
principle depend on $\mathbf{F}(s)$ by being a function of integrals of motion
(e.g.\ of the Hamiltonian or higher-derivative contour integrals):%
\begin{equation}
\alpha=\alpha\left(  \int\mathbf{F}^{\prime2}ds,\int\mathbf{F}^{\prime\prime
2}ds,\int K(s-\tilde{s})|\mathbf{F}(s)-\mathbf{F}(\tilde{s})|^{2}dsd\tilde
{s},\ldots\right)  \,. \label{dep}%
\end{equation}
The only requirement is that it should be invariant under the shifts of
parameter $\mathbf{F}(s)\rightarrow\mathbf{F}(s+h)$. Thus the calculation also
has to establish that $\alpha$ is a numerical constant, and more precisely
that $\alpha=\pi\mu^{2}$. In this case the Poisson bracket (\ref{bracket})
promotes upon quantization to precisely the conjectured commutator
(\ref{commutator}).

To conclude this section, we would like to note that in the
`bubbling AdS' case mentioned above, where the symplectic form was
computed using the direct CWZ method in \cite{MR,5auth}, the
consistency condition turns out to be much more powerful and in
fact fixes the symplectic form completely, i.e.~together with the
prefactor. The reason is that in the `bubbling AdS' case there is
non-trivial dynamics on the moduli space of geometries, namely the
planar droplets parametrizing it rotate with a particular angular
velocity, which can be fixed unambiguously by requiring that the
metric take the asymptotically AdS form with all perturbations
going to zero at the prescribed rate\footnote{The question whether
it is possible to make such an argument was first posed to me by
Boris Pioline.}.

\section{Fixing the prefactor}

\subsection{Simplifying assumptions}

In the previous section we used the consistency condition to show that the
time-independence of the D1-D5 solutions together with the form that the
Hamiltonian takes on the moduli space leave very little freedom for the
restricted symplectic form: it \textit{must} be given by Eq.~(\ref{toshow}),
and the only thing that remains is to evaluate the coefficient $\alpha$. This
is quite a strong restriction, since the most general expression for the
symplectic form respecting the translation invariance in the parameter space
is%
\begin{equation}
\Omega_{\text{general}}=\int ds\,d\tilde{s}\,K_{ij}(s-\tilde{s}|\mathbf{F}%
)\,\delta F_{i}(s)\wedge\delta F_{j}(\tilde{s})\,,
\end{equation}
where the \textit{symplectic kernel} $K$ has to be antisymmetric:%
\begin{equation}
K_{ij}(s|\mathbf{F})=-K_{ji}(-s|\mathbf{F})\,.
\end{equation}
Some extra restrictions could be generally derived using the invariance of
$\Omega$ under the Poincar\'{e} group acting on the curve, but these
restrictions would fall far short of what we have established: that the kernel
is diagonal in the perturbations and, most importantly, that it is
\textit{local}:%
\begin{equation}
K_{ij}=\frac{1}{2\alpha(\mathbf{F})}\delta_{ij}\delta^{\prime}(s)\,.
\label{local}%
\end{equation}
Although an explicit calculation will be necessary to determine $\alpha$, it
turns out that what we already know will allow us to organize this calculation
in a rather economical way.

\begin{figure}[t]
\centering {\ \includegraphics[width=4cm]{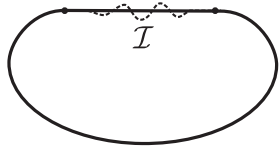} }\caption{The class of
closed curves containing a unit interval $\mathcal{I}$ along $x_{1}$, and
otherwise having an arbitrary profile. We evaluate the symplectic form around
any curve from this class, and for the perturbations supported on
$\mathcal{I}$ and directed along $x_{2}$. This symplectic form is then
extended to the full moduli space by uniqueness.}%
\label{curve}%
\end{figure}

Rather than evaluating the symplectic form in full generality, as it was done
in \cite{MR} for the `bubbling AdS' case, the idea is to find a case which is
simple yet restrictive enough so that it fixes the remaining freedom. We will
consider the following set of simplifying assumptions : 1) the curve contains
a straight-line interval $\mathcal{I}$ of the form (see Fig.~\ref{curve}):
\begin{equation}
F_{1}(s)=s,\quad F_{2,3,4}(s)=0\quad(0<s<1)\,; \label{curves}%
\end{equation}
2) the only nonzero component is%
\begin{equation}
\delta F_{2}(s)\equiv a(s)\quad(0<s<1)\,. \label{pert}%
\end{equation}

Under these assumptions, we will show that the symplectic form is given by%
\begin{equation}
\Omega=\frac{1}{2\pi\mu^{2}}\int a^{\prime}(s)\wedge a(s)\,ds\,. \label{Oms}%
\end{equation}
independently of the shape that the curve takes outside $\mathcal{I}$. A
moment's thought shows that this is enough to rule out any nontrivial
dependence of the kind (\ref{dep}). In other words, the only possible
expression of the form (\ref{toshow}),(\ref{dep}) which reduces to (\ref{Oms})
under the above simplifying assumptions is the one where $\alpha=\pi\mu^{2}$
is a numerical constant independent of $\mathbf{F}(s)$.

\subsection{Field perturbations}

Having assumed (\ref{curves}),(\ref{pert}), let us begin the evaluation of the
symplectic form. First of all we have to compute perturbations of the fields
$f_{1,5},A,B,\mathcal{C}$ in (\ref{micro}). It will be convenient to introduce
the polar coordinates in the space transverse to $\mathcal{I}$:%
\begin{align}
dx_{i}^{2}  &  =dx_{1}^{2}+dr^{2}+r^{2}(d\theta^{2}+\sin^{2}\theta\,d\phi
^{2}),\nonumber\\
r  &  \equiv\sqrt{x_{2}^{2}+x_{3}^{2}+x_{4}^{2}}\,,\quad x_{2}=r\cos
\theta\,\text{. }%
\end{align}
Starting with $f_{5}$, we can write its perturbed value as follows:%
\begin{align}
f_{5}(\mathbf{F}+\delta\mathbf{F})  &  =\bar{f}_{5}+\frac{Q_{5}}{L}%
\int_{-\infty}^{\infty}\frac{ds}{(x_{1}-s)^{2}+(x_{2}-a(s))^{2}+x_{3}%
^{2}+x_{4}^{2}}\nonumber\\
&  \approx\bar{f}_{5}+\frac{\pi Q_{5}}{Lr}+\frac{Q_{5}}{L}\int_{-\infty
}^{\infty}\frac{2x_{2}a(s)\,ds}{\left[  (x_{1}-s)^{2}+r^{2}\right]  ^{2}}\,,\\
\bar{f}_{5}  &  \equiv1+\frac{Q_{5}}{L}\int_{s\notin\mathcal{I}}ds\left(
\frac{1}{|\mathbf{x}-\mathbf{F}(s)|^{2}}-\frac{1}{(x_{1}-s)^{2}+r^{2}}\right)
\,, \label{fbar}%
\end{align}
where $\approx$ means that we expanded to the first order in $a(s)$. The
$\bar{f}_{5}$ is the regular part of $f_{5}$ (i.e. it has no singularities on
$\mathcal{I}$); it does not vary with $a(s).$ Using the momentum
representation for the convolution integral, we have%
\begin{align}
f_{5}  &  \approx\bar{f}_{5}+\gamma f^{\text{sing}},\quad\nonumber\\
f^{\text{sing}}  &  \equiv\frac{1}{r}+\frac{\cos\theta}{r^{2}}\,\left[
\left(  1+r|p|\right)  e^{-r|p|}\tilde{a}(p)\right]  ^{\vee}\,,\label{f5}\\
\gamma &  \equiv\frac{\pi Q_{5}}{L}\,=\frac{R}{2}.
\end{align}
Here we introduced the notation for the Fourier transform in $x_{1}$ and its
inverse:%
\begin{equation}
\tilde{a}(p)\equiv\int ds\,e^{ips}a(s),\qquad b(x)=\left[  \tilde
{b}(p)\right]  ^{\vee}\equiv\int\frac{dp}{2\pi}\,e^{-ipx_{1}}\tilde
{b}(p)\,.\nonumber
\end{equation}
Analogously we get%
\begin{align}
f_{1}  &  \approx\bar{f}_{1}+\gamma f^{^{\text{sing}}}\,,\label{f1}\\
A  &  \approx\bar{A}+\gamma f^{^{\text{sing}}}dx^{1}-\gamma\frac{\cos\theta
}{r}\left[  ipe^{-r|p|}\tilde{a}(p)\right]  ^{\vee}dx^{2}\,,
\end{align}
where the barred fields, whose precise form will not be needed below, are
again regular on $\mathcal{I}$. The $B$ and $\mathcal{C}$ are then found by
solving the flat-space Hodge dual equations in (\ref{micro}):%
\begin{align}
B\approx~  &  \bar{B}+\gamma\left\{  1-\cos\theta+\frac{\sin^{2}\theta}%
{r}\left[  \left(  1+r|p|\right)  e^{-r|p|}\tilde{a}(p)\right]  ^{\vee
}\right\}  d\phi\,,\\
\mathcal{C}\approx~  &  \bar{\mathcal{C}}+\gamma\left\{  -1+\cos\theta
-\frac{\sin^{2}\theta}{r}\left[  e^{-r|p|}\tilde{a}(p)\right]  ^{\vee
}\right\}  dx^{1}\wedge d\phi\nonumber\\
&  ~+~\gamma\cos\theta\sin\theta\left[  \left(  i\,\text{sign}\,p\right)
e^{-r|p|}\left(  2+r|p|\right)  \tilde{a}(p)\right]  ^{\vee}d\theta\wedge
d\phi\,. \label{C}%
\end{align}

\subsection{A coordinate transformation}

If we substitute (\ref{f5})-(\ref{C}) into (\ref{micro}) and expand in $a(s)$,
we will find what we call `naive' perturbations of $g_{\mu\nu},C_{\mu\nu}%
,\Phi$. Schematically we have:%
\begin{equation}
\delta g_{\mu\nu}^{\text{naive}}=g_{\mu\nu}(\mathbf{F}+\delta\mathbf{F}%
)-g_{\mu\nu}(\mathbf{F})\,
\end{equation}
expanded to the first order in $\delta\mathbf{F}$, with analogous expressions
for $\Phi$ and $C$. These perturbations will not yet be suitable for computing
the symplectic form using Eq.~(\ref{formIIB}). The reason is that $\delta
g_{\mu\nu}$ so defined will be singular on the curve in the coordinate system
in which $g_{\mu\nu}$\ is regular. A manifestation of this singularity is that
the $O(a(s))$ terms in (\ref{f5})-(\ref{C}) are all more singular as
$r\rightarrow0$ than the zero-order terms. The correct definition for the
field perturbations is (\cite{MR})%
\begin{equation}
\delta g_{\mu\nu}=g_{\mu\nu}^{(\varepsilon)}(\mathbf{F}+\delta\mathbf{F}%
)-g_{\mu\nu}(\mathbf{F})\,,
\end{equation}
where an appropriate change of coordinates $x^{\mu}\rightarrow x^{\mu
}-\varepsilon^{\mu}$ has to be applied to $g_{\mu\nu}(\mathbf{F}%
+\delta\mathbf{F})$ before the subtraction is made. The effect of this can be
expressed as
\begin{equation}
\delta g_{\mu\nu}=\delta g_{\mu\nu}^{\text{naive}}+\varepsilon^{\lambda}%
g_{\mu\nu,\lambda}+2\varepsilon_{,(\mu}^{\lambda}g_{\nu)\lambda}\,, \label{v1}%
\end{equation}
where the additional terms (which can also be written as $\nabla_{(\mu
}\varepsilon_{\nu)}$) are the effect of an $O(a)$ coordinate transformation.
They have to be chosen so that the resulting $\delta g_{\mu\nu}$ be regular.
Analogously we will have%
\begin{align}
\delta\Phi &  =\delta\Phi^{\text{naive}}+\varepsilon^{\lambda}\Phi
_{\,,\lambda}\,,\label{v2}\\
\delta C_{\mu\nu}  &  =\delta C_{\mu\nu}^{\text{naive}}+\varepsilon^{\lambda
}C_{\mu\nu,\lambda}+2\varepsilon_{,(\mu}^{\lambda}C_{\nu)\lambda}%
+\Lambda_{\lbrack\mu,\nu]}\,, \label{v3}%
\end{align}
where in the case of the two-form field we also have to include the effect of
an abelian gauge transformation.

\begin{figure}[t]
\centering {\ \includegraphics[width=4cm]{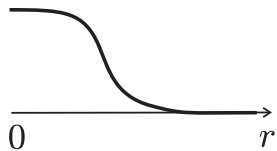} }\caption{The `cutoff'
$C^{\infty}$ function $\chi(r)$ interpolates smoothly between 1 at $r=0$ and 0
at $r=\infty$.}%
\label{chi}%
\end{figure}

It is not difficult to guess the form of a coordinate
transformation $x\rightarrow\tilde{x}$ needed to make the field
perturbations regular. The effect of this transformation should be
such that 1) the perturbed curve $x_{2}=a(x_{1})$ has equation
$\tilde{x}_{2}=0$ in the new coordinates; 2) the transformation
has unit Jacobian on the curve; 3) the transformation tends to the
identity transformation at infinity. The following transformation
satisfies all these requirements and will do the job:
\begin{align}
\tilde{x}_{1} &  =x_{1}+a^{\prime}(x_{1})x_{2}\chi(r)\,,\nonumber\\
\tilde{x}_{2} &  =x_{2}-a(x_{1})\chi(r)\,.\label{trans}%
\end{align}
Here $\chi(r)$ is any function which interpolates smoothly between 1 at $r=0$
and 0 at $r=\infty$ (see Fig.\ \ref{chi}). Thus the only nonzero components of
$\varepsilon^{\mu}$ are
\begin{equation}
\varepsilon^{1}=-a^{\prime}(x_{1})x_{2}\chi(r),\quad\varepsilon^{2}%
=a(x_{1})\chi(r)\,.
\end{equation}

\subsection{Reduction to the plane wave}

Implementing this coordinate transformation, we will get regular
field variations, which could in principle be used to compute the
symplectic form using Eq~(\ref{formIIB}). Note that the barred
terms in (\ref{f5})-(\ref{C}) will `contaminate' the field
variations as a consequence of expansions that will have to be
done once (\ref{f5})-(\ref{C}) is substituted into (\ref{micro}),
and later via the second terms in (\ref{v1})-(\ref{v3}). However,
we claim that all these contaminating contributions can be
ignored. More precisely, we would like to argue that the
symplectic form can be computed by putting all the barred terms to
zero both in the unperturbed spacetime and in the perturbations.

The argument why this is so consists of three steps. First of all, we notice
that to determine the \textit{local} contribution to the symplectic current,
it is enough to perform the integration in (\ref{cauchy}) in a small
neighborhood of $\mathcal{I}$, which should include the regions where
$\chi\neq0$. The reason is that outside of this neighborhood the field
perturbations depend on $a(s)$ in a \textit{regular nonlocal} way. Thus the
same will be true of the symplectic current, and upon integration this region
can only give a \textit{nonlocal} contribution to the symplectic form. Such a
contribution cannot affect the coefficient in front of the \textit{local}
symplectic kernel (\ref{local}), which we have to determine.

Second, once we have limited ourselves to a small neighborhood of
$\mathcal{I},$ in this region we can expand the barred fields in the
transverse space. Schematically we can write:%
\begin{equation}
\bar{f}=\bar{f}|_{r=0}+x_{\perp}^{i}\partial_{i}\bar{f}|_{r=0}+\ldots,\quad
x_{\perp}^{i}=(x_{2},x_{3},x_{4})\,, \label{coeff}%
\end{equation}
where $\bar{f}$ stands for any of the barred fields that we
encountered. The coefficients in this series are fields of
$x_{1}=s$ only. Now, it is easy to check that the metric and the
perturbations will depend on these coefficients analytically near
$\mathcal{I}$. In particular, the limit when all the barred fields
tend to zero is completely well behaved. Away from this limit, the
coefficients of (\ref{coeff}), say $\bar{f}(s)$, will enter the
field perturbations as functions multiplying $a(s)$,
$a^{\prime}(s)$, or general convolution integrals of the form%
\begin{equation}
\int a(s-\tilde{s})\Psi\left(  \frac{\tilde{s}}{r}\right)  d\tilde{s}\,,
\end{equation}
evaluated \textit{at the same point} $s$. It means that after
integration, these coefficients may make their way into the
symplectic form only via terms involving products $\bar{f}(s)a(s)$
and such.

As a third, final, step, we notice that there is a lot of freedom to deform
the curve outside $\mathcal{I}$ to make the barred fields into functions
varying nontrivially along and near $\mathcal{I}$. This can be seen by
examining the explicit expression (\ref{fbar}) for $\bar{f}_{5}$ and analogous
expressions for the other fields. However, for varying $\bar{f}$, the terms
involving products like $\bar{f}(s)a(s)$ would not be consistent with the
translationally invariant nature of the symplectic kernel (\ref{local}). This
means that such terms simply cannot appear, which finishes the argument.

Putting the barred fields to zero means that we pass from the
closed curve to an infinite straight line and also drop the 1's
from $f_{1,5}$. The resulting nonperturbed geometry is a plane
wave. In other words we proved that to find the general symplectic
form it is enough to find the symplectic form around the plane
wave.

\subsection{Plane wave symplectic form}

In this subsection we evaluate the symplectic form for perturbations around
the plane wave, which corresponds to putting all the barred fields in
(\ref{f5})-(\ref{C}) to zero. At this point bulky computations are
unavoidable, however they lend themselves easily to automatization. The reader
may want to skip to the main results: the gravitational and~the two-form
symplectic current integrals (\ref{JG}),(\ref{JF}), and the symplectic form
(\ref{Om}),(\ref{Omega}) which agrees with (\ref{Oms}). According to the
arguments of the preceding subsections, this completes the proof that the
symplectic form on the full moduli space is given by Eq.~(\ref{toshow}) with
the right prefactor.

Notice that once we dropped the barred fields we have $f_{1}=f_{5}$ to the
first order in $a$. This means that the dilaton stays constant: $\Phi=1$,
$\delta\Phi=0$. The difference between the Einstein and string frame
disappears. Also the $T^{4}$ part of the metric becomes trivial and just
contributes a volume factor to the symplectic form upon integration. The
unperturbed 6d metric is:%
\begin{equation}
ds^{2}=-\frac{r}{\gamma}dt^{2}-dt\,dx_{1}+\frac{\gamma}{r}dr^{2}+r\left[
\gamma d\theta^{2}+2(1-\cos\theta)(\gamma d\phi^{2}-d\phi\,dy)+\frac{dy^{2}%
}{\gamma}\right]  \,.
\end{equation}
The radial part of the metric becomes regular in the variable
$\rho=\sqrt{r}$. To see that the angular part is regular, one
should notice that near $\theta=\pi$ the $(\phi,y)$ part rewrites
neatly as $\gamma(2d\phi+\gamma ^{-1}dy)^{2}$. In fact the metric
of the transverse space can be brought to the flat
$\mathbb{R}^{4}$ form, but we will find it convenient to keep
working in the current coordinate system.

The pertubartions of the 6d metric after the coordinate transformation
(\ref{trans}) become (the Fourier transform $\widetilde{\delta g}_{\mu\nu}$ is
equal $\tilde{a}(p)$ times the coefficient given in the table):%
\begin{align}
tt &  :\frac{\cos\theta}{\gamma}\left[  (1+r|p|)e^{-r|p|}-\chi\right]  \nonumber\\
tx_{1} &  :-\cos\theta\,r\chi\,p^{2}\nonumber\\
tr &  :\cos\theta\left(  e^{-r|p|}-\chi-r\chi^{\prime}\right)  ip\nonumber\\
t\theta &  :-\sin\theta\,r\left(  e^{-r|p|}-\chi\right)  ip\nonumber\\
x_{1}r &  :\frac{\gamma\cos\theta\,}{r}\left(  e^{-r|p|}-\chi\right)  ip\nonumber\\
x_{1}\theta &  :-\gamma\sin\theta\,\left(  e^{-r|p|}-\chi\right)
ip \nonumber\\
rr &  :\frac{\gamma\cos\theta}{r^{2}}\,\left[
(1+r|p|)e^{-r|p|}-\chi
+2r\chi^{\prime}\right]  \nonumber\\
r\theta &  :-\gamma\sin\theta\,\chi^{\prime}\nonumber\\
\theta\theta &  :\gamma\cos\theta\,\left[  (1+r|p|)e^{-r|p|}-\chi\right]
\nonumber\\
\phi\phi &  :2\gamma(1-\cos\theta)\left[  (1+r|p|)e^{-r|p|}-\chi\right]
\nonumber\\
\phi y &  :(1-\cos\theta)\left[  (1+r|p|)e^{-r|p|}-\chi\right]  \nonumber\\
yy &  :-\frac{\cos\theta}{\gamma}\left[  (1+r|p|)e^{-r|p|}-\chi\right]
\end{align}
We see the effect that the presence of $\chi$ has on the regularity on the
field perturbations: it smoothens the behavior at $r=0$ by subtracting the
leading singularity$.$ A quick check of the regularity at $\theta=\pi$ for all
$r$ can be performed by noticing that the perturbation near the south pole can
be written as%
\begin{equation}
\gamma(2d\phi+\gamma^{-1}dy)^{2}\left[  (1+r|p|)e^{-r|p|}-\chi\right]  \,.
\end{equation}
To compute the symplectic form, we choose the Cauchy surface $\Sigma
=\{t=const\},$ so that the only necessary symplectic current component is
$J^{t}$. The gravitational symplectic current evaluated using the
Crnkovi\'{c}-Witten formula (\ref{CW}) and the above metric perturbations is
equal%
\begin{align}
J_{g}^{t}(r,\theta) &  =\frac{\gamma\sin\theta}{r}\int\frac{dp}{2\pi
}K(p,r,\theta)\,ip\,\tilde{a}(p)\wedge\tilde{a}(-p)\,,\nonumber\\
K &  =e^{-2r|p|}\left[  \cos^{2}\theta\left(  3+4r|p|+2r^{2}p^{2}\right)
-1-2r|p|\right]  \nonumber\\
&  \quad+e^{-r|p|}\Bigl\{\chi\left[  \cos^{2}\theta\left(  r^{3}%
|p|^{3}-4r|p|-6\right)  +2+2r|p|-2r^{2}p^{2}\right]  \nonumber\\
&  \quad\quad+\chi^{\prime}r\left[  \cos^{2}\theta\left(  4+2r|p|-r^{2}%
p^{2}\right)  +r|p|-1\right]  -\chi^{\prime\prime}r^{2}\cos^{2}\theta
\Bigr\}\nonumber\\
&  \quad+\cos^{2}\theta\,\left[  \chi^{2}\left(  3-r^{2}p^{2}\right)
+\chi\chi^{\prime}r\left(  2r^{2}p^{2}-4\right)  +\chi^{\prime2}r^{2}+\chi
\chi^{\prime\prime}r^{2}\right]  \nonumber\\
&  \quad\quad+\chi^{2}(r^{2}p^{2}-1)+\chi\chi^{\prime}r\,.
\end{align}
We see that $K=O(r)$ for $r\rightarrow0$ due to the presence of $\chi$, and
thus the symplectic current is regular. Integrating in $\theta$, we get%
\begin{align}
\int d\theta\,J_{g}^{t}(r,\theta) &  =\frac{2}{3}\gamma\int\frac{dp}{2\pi
}\left[  k_{1}+k_{2}+k_{3}\right]  \,ip\,\tilde{a}(p)\wedge\tilde
{a}(-p)\,,\nonumber\\
k_{1} &  =e^{-2r|p|}(2r|p|-2)|p|\,,\nonumber\\
k_{2} &  =e^{-r|p|}\left[  \chi|p|(2-6r|p|+r^{2}p^{2})+\chi^{\prime}\left(
1+5r|p|-r^{2}p^{2}\right)  -\chi^{\prime\prime}r\right]  \,,\nonumber\\
k_{3} &  =2\chi^{2}rp^{2}+\chi\chi^{\prime}\left(  2p^{2}r^{2}-1\right)
+r\chi^{\prime2}+r\chi\chi^{\prime\prime}\,.
\end{align}
Now it remains to integrate in $r$. We have%
\begin{equation}
\int dr\,k_{1}=-1/2\,.
\end{equation}
On the other hand, a couple of integrations by part show that%
\begin{align}
\int dr\,k_{2} &  =-2\chi|_{r=0}=-2\,,\\
\int dr\,k_{3} &  =\chi^{2}|_{r=0}=1\,.
\end{align}
Thus the total value
\begin{equation}
\int dr\,d\theta\,J_{G}^{t}(r,\theta)=-\gamma\int\frac{dp}{2\pi}%
\,ip\,\tilde{a}(p)\wedge\tilde{a}(-p)\,.\label{JG}%
\end{equation}
Now let's consider the two form. The unperturbed value:
\begin{equation}
C=r\left(  1-\cos\theta\right)  dt\wedge d\phi+\frac{r}{\gamma}dt\wedge dy\,.
\end{equation}
Again to study the unperturbed geometry one could in principle go
to the regular coordinates in which the transverse metric is
$dX_{i}^{2}$ and the three-form field strength would become
$\propto dt\wedge\left(  dX_{1}\wedge dX_{2}+dX_{3}\wedge
dX_{4}\right)  $. However, in the problem at hand such a change is
unnecessary. We simply compute the two-form perturbations using
(\ref{f5})-(\ref{C}) in (\ref{micro}) (with all the barred fields
put to zero) and applying the coordinate change (\ref{trans}).
They still come out singular
near the south pole. An abelian gauge transformation with the paramater%
\begin{equation}
\Lambda=\left[  r\chi\cos\theta\,ip\tilde{a}(p)\right]  ^{\vee}dy
\end{equation}
has to be applied. After that the two-form perturbations become (the Fourier
transforms of the corresponding components are equal $\tilde{a}(p)$ times the
coefficient in the table):%
\begin{align}
dt\wedge d\phi &  :\left(  1-\cos\theta\right)  \left[  \left(  1+r|p|\right)
e^{-r|p|}-\chi\right]  \nonumber\\
dx_{1}\wedge d\phi &  :\gamma\sin^{2}\theta\,e^{-r|p|}|p|\nonumber\\
dr\wedge d\phi &  :\gamma(\cos^{2}\theta-\cos\theta)e^{-r|p|}ip\nonumber\\
d\theta\wedge d\phi &  :\gamma\sin\theta(r|p|+2\cos\theta)e^{-r|p|}%
i\,\text{sign}\,p\nonumber\\
dt\wedge dy &  :-\frac{\cos\theta}{\gamma}\left[  \left(  1+r|p|\right)
e^{-r|p|}-\chi\right]  \nonumber\\
dr\wedge dy &  :-\cos\theta\,e^{-r|p|}ip\nonumber\\
d\theta\wedge dy &  :\sin\theta\,r\,e^{-r|p|}ip
\end{align}
This two-form is regular, and a quick way to check this is to see
that near $\theta=0$ and $\theta=\pi$ it looks like
$dy\wedge(\ldots)$ and $(2d\phi+\gamma^{-1}dy)\wedge(\ldots)$,
respectively.

Now we can compute the two-form symplectic current using Eq.~(\ref{Fs}), which
comes out to be%
\begin{align}
J_{F}^{t}(r,\theta) &  =-\gamma r\sin\theta\cos^{2}\theta\int\frac{dp}{2\pi
}e^{-2r|p|}(2r|p|+2)i|p|^{3}\,\tilde{a}(p)\wedge\tilde{a}(-p)\,,\\
\int dr\,d\theta\,J_{F}^{t}(r,\theta) &  =-\gamma\int\frac{dp}{2\pi
}ip\,\,\tilde{a}(p)\wedge\tilde{a}(-p)\,.\label{JF}%
\end{align}
We see that, unlike for the metric, in this case the $\chi$ regulator does not
make its way into the symplectic current.

The total symplectic form is obtained by adding the contributions of the
metric and the two-form:%
\begin{equation}
\Omega=\frac{1}{(2\pi)^{7}g_{s}^{2}}(2\pi)(2\pi R)(2\pi)^{4}V_{4}\int
dr\,d\theta\,(J_{G}^{t}+J_{F}^{t})\,,\label{Om}%
\end{equation}
where we took into account volume factors appearing because of the integration
over the $\phi$ and the $y$ circles, and over $T^{4}$. All in all, the final
result is%
\begin{equation}
\Omega=-\frac{1}{2\pi\mu^{2}}\int\frac{dp}{2\pi}ip\,\,\tilde{a}(p)\wedge
\tilde{a}(-p)\,,\label{Omega}%
\end{equation}
which is equivalent to (\ref{Oms}). Q.E.D.

\section{Discussion}

In this paper we have quantized the moduli space of the regular D1-D5
geometries, and used this result to count the D1-D5 black hole microstates,
directly from SUGRA. The method is based on our previous work \cite{MR,5auth},
and incorporates a new ingredient---the consistency condition---which allowed
us to reduce the computational work needed to obtain the general result. Most
certainly, this new ingredient will play a role in the future applications of
the symplectic form quantization to the other SUGRA moduli spaces, such as the
3-charge moduli space parametrized by hyper-Kaehler manifolds \cite{BW}.

Our result fits nicely with the general idea of Mathur's program
\cite{Mathur}, which can be loosely described as aiming to understand the
inner structure of quantum black holes purely from SUGRA, i.e.~without
D-branes. Recently, several interesting results breathed new life into the
program and opened up new avenues of research. In \cite{Vijay}, it was
demonstrated how the emergence of an effective black hole geometry may be seen
from AdS/CFT correlators in a typical D1-D5 ground state. In \cite{Jan}, it
was shown how an effective geometry may be recovered by studying boundary
1-point functions. It would be very interesting to combine these two
approaches and derive a truly emergent horizon of the D1-D5 black hole.
Another problem which needs to eventually be addressed is the derivation of
the nonzero area horizon from the curvature-corrected SUGRA Lagrangian. In the
best of the worlds, these two derivations would produced horizons of the same
size. The future will show if this is indeed the case.

\acknowledgments

 I would like to thank V. Balasubramanian, J. de
Boer, T. Jacobson, T. Levi, O.~Lunin, D.~Marolf, D. Martelli, A.
Naqvi, P. van Nieuwenhuizen, M. Porrati, R. Schiappa, J. Simon and
J.-T. Yee for useful comments and discussions. I would like to
thank A. Shomer for the early collaboration. I would like to
especially thank Liat Maoz for the fruitful collaboration on the
quantization of SUGRA solutions, the early collaboration on this
project, and for the many invaluable discussions. The results of
this paper were first reported on October 25, 2005, at the
Algebraic Geometry and Topological Strings conference in the
Instituto Superior T\'{e}cnico Lisbon, and in early November 2005
at the theory group seminars of the SUNY Stony Brook, the NYU, the
University of Maryland, and the University of Pennsylvania. I
would like to thank all these institutions for their hospitality.
This work is supported by the EU under RTN contract
MRTN-CT-2004-503369.


\begin{thebibliography}{99}                                                                                               %


\bibitem {Mathur}As reviewed e.g.~in S.~D.~Mathur, \textquotedblleft The
fuzzball proposal for black holes: An elementary review,\textquotedblright%
\ Fortsch.\ Phys.\ \textbf{53}, 793 (2005) [arXiv:hep-th/0502050].


\bibitem {MR}L.~Maoz and V.~S.~Rychkov, ``Geometry quantization from
supergravity: The case of 'bubbling AdS','' JHEP \textbf{0508}, 096 (2005)
[arXiv:hep-th/0508059].


\bibitem {5auth}L.~Grant, L.~Maoz, J.~Marsano, K.~Papadodimas and
V.~S.~Rychkov, ``Minisuperspace quantization of 'bubbling AdS' and free
fermion droplets,'' JHEP \textbf{0508}, 025 (2005) [arXiv:hep-th/0505079].

\bibitem {LM}
O.~Lunin and S.~D.~Mathur, ``Metric of the multiply wound rotating
string,'' Nucl.\ Phys.\ B \textbf{610}, 49 (2001)
[arXiv:hep-th/0105136];
\newline
O.~Lunin and S.~D.~Mathur,
  ``AdS/CFT duality and the black hole information paradox,''
  Nucl.\ Phys.\ B {\bf 623}, 342 (2002)
  [arXiv:hep-th/0109154].

\bibitem {LMM}O.~Lunin, J.~Maldacena and L.~Maoz, \textquotedblleft Gravity
solutions for the D1-D5 system with angular momentum,\textquotedblright%
\ arXiv:hep-th/0212210.

\bibitem{Marolf} B.~C.~Palmer and D.~Marolf, \textquotedblleft Counting
supertubes,\textquotedblright\ JHEP \textbf{0406}, 028 (2004)
[arXiv:hep-th/0403025]; 
\newline
  D.~Bak, Y.~Hyakutake, S.~Kim and N.~Ohta,
  ``A geometric look on the microstates of supertubes,''
  Nucl.\ Phys.\ B {\bf 712}, 115 (2005)
  [arXiv:hep-th/0407253];

\bibitem {Sen}The following papers are meant to provide a point of entry to
the massive recent literature: \newline A.~Sen, \textquotedblleft How does a
fundamental string stretch its horizon?\textquotedblright\ JHEP \textbf{0505},
059 (2005) [arXiv:hep-th/0411255];
\newline V.~Hubeny, A.~Maloney and M.~Rangamani, \textquotedblleft
String-corrected black holes,\textquotedblright\ JHEP \textbf{0505}, 035
(2005) [arXiv:hep-th/0411272].


\bibitem {Marika}M.~Taylor, \textquotedblleft General 2 charge
geometries,\textquotedblright\ arXiv:hep-th/0507223.


\bibitem {BW}See O.~Lunin,
  ``Adding momentum to D1-D5 system,''
  JHEP {\bf 0404}, 054 (2004)
  [arXiv:hep-th/0404006];
  \newline
I.~Bena and N.~P.~Warner, \textquotedblleft Bubbling supertubes
and foaming black holes,\textquotedblright\ arXiv:hep-th/0505166;
\newline P.~Berglund, E.~G.~Gimon and T.~S.~Levi, \textquotedblleft
Supergravity microstates for BPS black holes and black
rings,\textquotedblright\ arXiv:hep-th/0505167,
\newline as well as references therein for the previous work on the 3-charge geometries.

\bibitem {CW}\v{C}.~Crnkovi\'{c} and E.~Witten, \textquotedblleft Covariant
Description Of Canonical Formalism In Geometrical Theories,\textquotedblright%
\ in \emph{Three hundred years of gravitation}, Eds.\ S.W.~Hawking and
W.~Israel (Cambridge University Press, 1987), p.676.


\bibitem {Z}G.~J.~Zuckerman, \textquotedblleft Action Principles And Global
Geometry," in \emph{Mathematical Aspects Of String Theory}, San Diego 1986,
Proceedings, Ed.\ S.-T.~Yau (Worls Scientific, 1987), p.259.


\bibitem {LLM}H.~Lin, O.~Lunin and J.~Maldacena, \textquotedblleft Bubbling
AdS space and 1/2 BPS geometries,\textquotedblright\ JHEP \textbf{0410}, 025
(2004) [arXiv:hep-th/0409174].


\bibitem {Vijay}V.~Balasubramanian, P.~Kraus and M.~Shigemori,
\textquotedblleft Massless black holes and black rings as effective geometries
of the D1-D5 system,\textquotedblright\ Class.\ Quant.\ Grav.\ \textbf{22},
4803 (2005) [arXiv:hep-th/0508110].


\bibitem {Jan}L.~F.~Alday, J.~de Boer and I.~Messamah, ``What is the dual of a
dipole?,'' arXiv:hep-th/0511246.


\bibitem {Jevicki}A.~Donos and A.~Jevicki, \textquotedblleft Dynamics of
chiral primaries in $AdS_{3}\times S^{3}\times T^{4}$,\textquotedblright%
\ arXiv:hep-th/0512017.

\end{thebibliography}
\end{document}